\documentclass{webofc}

\usepackage[varg]{txfonts}   
\usepackage{hyperref}
\usepackage{url}
\usepackage[export]{adjustbox}
\hypersetup{colorlinks=true,citecolor=blue,urlcolor=blue,linkcolor=blue}
\begin{document}
\title{QCD phase structure \& equation of state:\\ A functional perspective\thanks{Invited plenary talk at the 31st International Conference on Ultra-relativistic Nucleus-Nucleus Collisions (Quark Matter 2025), 
Frankfurt, Germany, Apr 6-12, 2025.}}
%
%

\author{\firstname{Fabian} \lastname{Rennecke}\inst{1,2}\fnsep\thanks{\email{fabian.rennecke@theo.physik.uni-giessen.de}}
}

\institute{Institute for Theoretical Physics, Justus Liebig University Giessen, 35392 Giessen, Germany 
\and
Helmholtz Research Academy Hesse for FAIR (HFHF), Campus Giessen, 35392 Giessen, Germany
          }

\abstract{The phase structure of QCD remains an open fundamental problem of standard model physics. In particular at finite density,
our knowledge is limited. Yet, numerous model studies point towards a rich and complex phase diagram at large density. 
Functional methods like the functional renormalization group and Dyson-Schwinger equations offer a way to study hot and dense QCD matter directly from first principles. I will discuss the phase structure of QCD and its experimental signatures through the lens of these methods.
}
\maketitle
\vspace{-5pt}
\section{Introduction}\label{sec:intro}
\vspace{-5pt}

The direct calculation of the QCD phase diagram is an inherently challenging task for three key reasons. First, phase transitions are associated with the formation of condensates through resonant interactions. This is, in almost all cases, a strong coupling effect and hence nonperturbative.
Weak-coupling results can only cover the "edges" of the phase diagram,
while most of the phenomenologically relevant regions, especially in the context of heavy-ion collisions and neutrons stars, are nonperturbative. Based on models and exploratory QCD studies, numerous exotic phases are possible here \cite{Fukushima:2010bq}.
Second, different phases are dominated by different effective degrees of freedom. The construction of an effective field theory that describes the whole phase diagram is hence impossible, as this
impedes systematic power counting. Third, the phenomenologically relevant cases of, for example, nonzero baryon chemical potential, $\mu_B$, and real time lead to sign problems, severely limiting the applicability of any importance sampling-based method like conventional lattice QCD.

However, owing to rapid progress in recent years, first direct results on the phase diagram at nonzero $\mu_B$ from first principles are available from functional methods, i.e.\ Dyson Schwinger equations (DSEs) and the functional renormalization group (FRG). In these proceedings, I highlight this progress in broad strokes.

\vspace{-5pt}
\section{Functional methods}\label{sec:func}
\vspace{-5pt}

Functional methods are, as the name suggests, based on the functional formulation of quantum field theory (QFT). The basic object is the path integral (the vacuum amplitude),
\begin{align}\label{eq:Z}
Z[J] = \int\!\! \mathcal{D}\Phi\, e^{iS[\Phi]+i\int_xJ(x)\Phi(x)}\,,\qquad \big\langle \Phi(x_1) \cdots\Phi(x_n)\big\rangle \sim \frac{\delta}{\delta J(x_1)} \cdots \frac{\delta}{\delta J(x_n)}\, Z[J] \bigg|_{J=0}\,,
\end{align}
where $\mathcal{D}\Phi$ denotes an integral over all possible configurations of quantum fields $\Phi(x)$. One may think of the exponential in the path integral as a distribution, and the correlation functions, 
which are obtained from functional derivatives with respect to the sources $J(x)$, as moments of this distribution. Hence, solving a QFT is equivalent to knowing all correlation functions. DSE and FRG provide exact relations for these correlation functions, and are therefore exact, but usually more practical, reformulations of the original path integral. By specifying the classical action $S[\Phi]$ in Eq.\ (\ref{eq:Z}), one chooses the theory/model one wishes to study. So for QCD one chooses the QCD action in terms of gluons and quarks, $\Phi = \{A_\mu,q,\bar q\}$.

DSEs are the generalization of the classical Euler-Lagrange equations to QFT. FRG is based on gradually including fluctuations of larger size by successively integrating out field modes with increasingly small momenta. In both cases, Eq.\ (\ref{eq:Z}) is converted into a set of coupled differential equations. There are infinitely many correlations functions, giving rise to infinite towers of these equations. Truncations are clearly necessary. This does not mean that it is assumed that some of these correlations are zero, only that their feedback into the quantities of interest is small. Except for weak coupling and, in case of the FRG, near second-order phase transitions \cite{Balog:2019rrg}, small parameters to guide these truncations are unknown. However, a hierarchy from low to high-order correlations is typically observed, facilitating systematic studies \cite{Ihssen:2024miv, Huber:2025kwy}. Still, systematic error control is the biggest challenge for functional methods. For QCD-related reviews, see \cite{Fischer:2018sdj, Fu:2022gou} and references therein.

\begin{figure}[t]
\centering
\sidecaption
\includegraphics[width=6.9cm,clip]{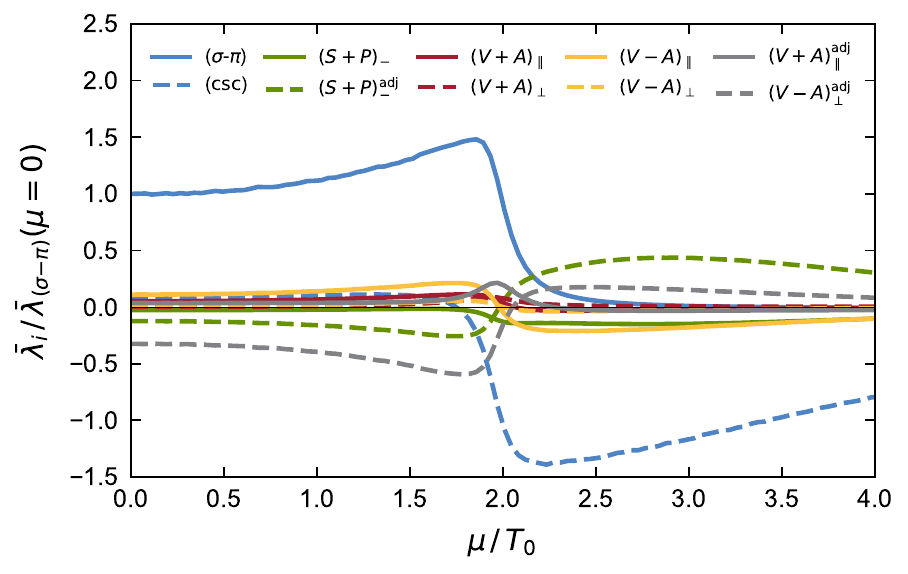}
\caption{Complete set of pointlike four-quark interaction channels for two-flavor QCD as functions of the quark chemical potential $\mu$ at temperatures $T_0(\mu)$ just above the transition temperature from the FRG \cite{Braun:2019aow}. The dominance of the $(\sigma\!-\!\pi)$-channel at small $\mu$ indicates chiral symmetry breaking, while the dominance of the (csc)-channel shows that QCD prefers to be in a color-superconducting state at large $\mu$.}
\label{fig:chan}  
\vspace{-8pt}   
\end{figure}

In addition to the absence of sign problems, bound states and condensates naturally emerge from the underlying elementary correlations of quarks and gluons in functional QCD. These structures appear as resonances in quark and gluon scattering channels which, in turn, can be constructed systematically from complete sets of basis tensors. This allows for an unbiased determination of the phase structure. An example is shown in Fig.\ \ref{fig:chan}.

\vspace{-5pt}
\section{Phase transitions}\label{sec:phase}
\vspace{-5pt}

\begin{figure}[t]
\centering
\sidecaption
\includegraphics[width=7.0cm,clip]{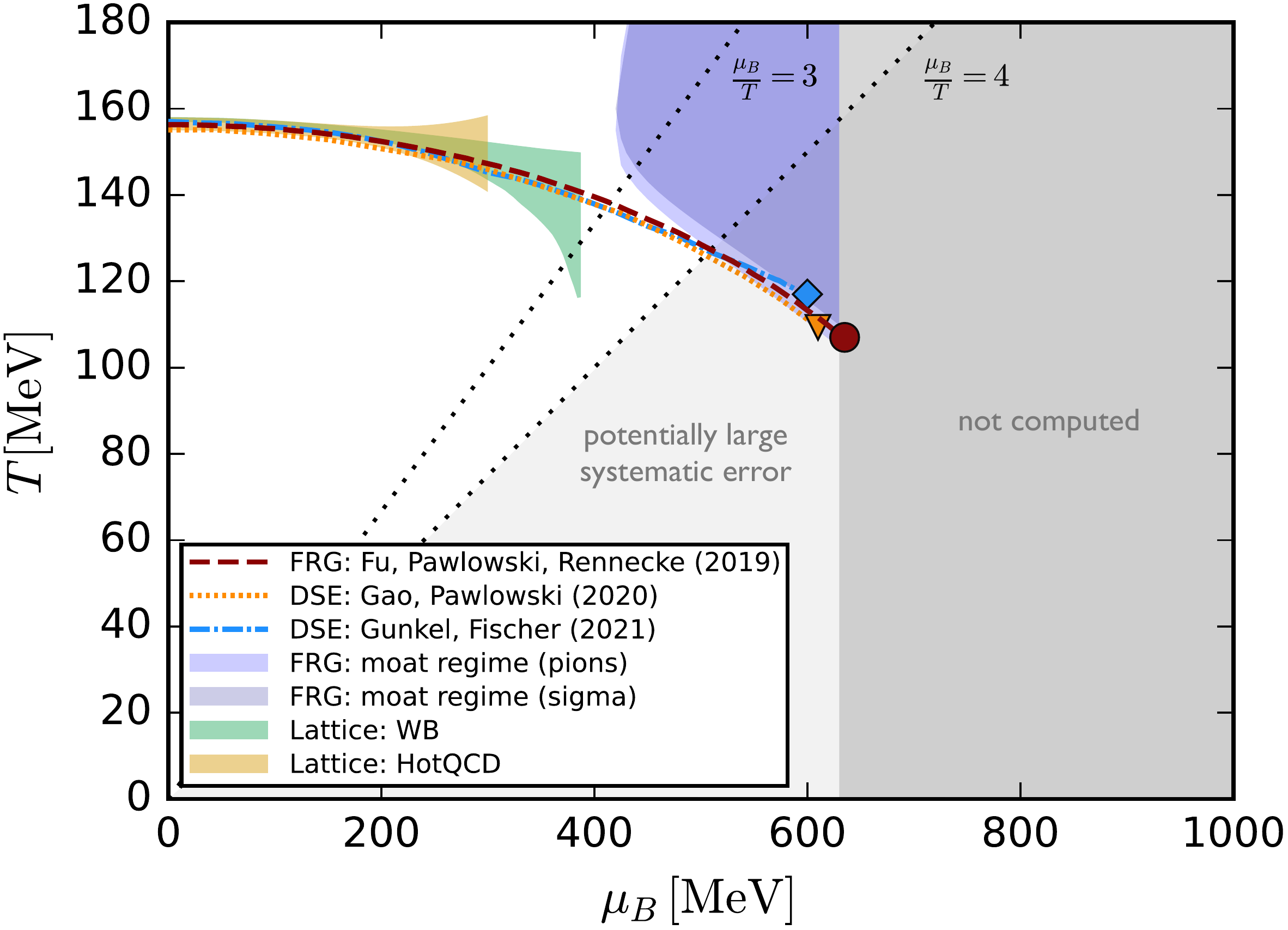}
\caption{The QCD phase diagram from functional methods. The broken lines are the chiral crossovers obtained directly from functional QCD. For comparison, the bands show extrapolations from lattice QCD. These lines end in a CEP and, within the underlying approximations, continue as first-order transitions. However, the systematic error may be large in the gray region, so this is not shown. The blue region is the moat regime, see Sec.\ \ref{sec:moat}.}
\label{fig:PD} 
\vspace{-8pt}   
\end{figure}

A natural starting point for the investigation of the phase diagram is the extension of the results from lattice QCD at zero $\mu_B$ \cite{Bellwied:2015rza, Bazavov:2018mes} to nonzero $\mu_B$. To this end, truncated sets of DSEs and FRG equations for the chiral condensate $\langle \bar q q \rangle$ need to be solved. In this case, the available lattice QCD results can serve as a benchmark for the quality of the underlying approximations at small $\mu_B$. The first functional QCD result in line with this benchmark were obtained from the FRG in \cite{Fu:2019hdw} and DSEs in \cite{Gao:2020fbl, Gunkel:2021oya}. The resulting phase diagram is shown in Fig.\ \ref{fig:PD}. While the systematic error has originally been estimated to be potentially large at $\mu_B/T \gtrsim 4$ for the different functional QCD results, they show remarkable agreement in particular regarding the location of the critical endpoint (CEP). Together with various published (e.g., \cite{Lu:2025cls}) and forthcoming systematic improvements, this points at a CEP location of $(T,\mu_B)_{\rm CEP} \approx (110,630)\,{\rm MeV}$ with an uncertainty of about 10\%. This result is corroborated by subsequent extrapolations of lattice QCD results based on Yang-Lee edge singularities \cite{Basar:2023nkp, Clarke:2024ugt} and thermodynamic relations \cite{Shah:2024img}.

All these results come with a caveat: they were all obtained under the assumption of a homogeneous phase, i.e., with a uniform order parameter. At this point, it can therefore not be excluded that there is a, possibly inhomogeneous, state of QCD at $\mu_B\gtrsim 600$\,MeV with a lower free energy than any of the homogeneous states that have been explored so far. This is at least suggested also by the presence of the moat regime in this region, cf.\ Sec.\ \ref{sec:moat}. To be more cautious, as of now, we can only say that the homogeneous chiral crossover is very likely to end around $(T,\mu_B) \approx (110,630)\,{\rm MeV}$. Whether it ends in a CEP, some spatially modulated phase, or both, is not certain yet.

\begin{figure}[b]
\centering
\includegraphics[width=6cm,clip,valign=t]{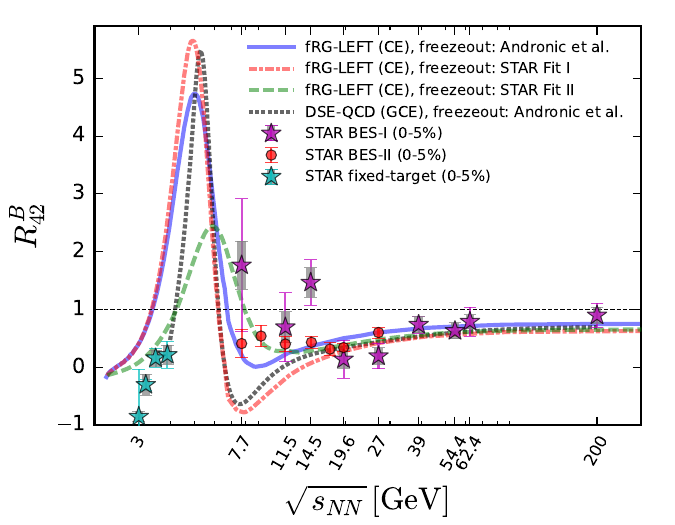}
\hfill
\includegraphics[width=6.5cm,clip,valign=t]{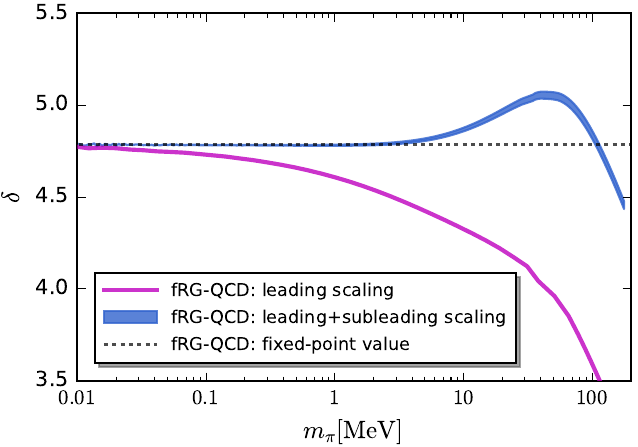}
\caption{\emph{Left:} Kurtosis of the net-baryon distribution from functional methods \cite{Fu:2023lcm, Lu:2025cls} together with the kurtosis of net-protons measured by STAR \cite{Lu:2025cls}. \emph{Right:} Determination of the size of the chiral $O(4)$ scaling region though the critical exponent $\delta$ \cite{Braun:2023qak}.}
\label{fig:crit} 
\vspace{-8pt}    
\end{figure}

It is imperative to support the theoretical search for the CEP and other exotic structures with experiments such as heavy-ion collisions. A key feature of a CEP, and any second-order phase transition, is that in its vicinity universal critical phenomena determine main features of the system. This can, for example, lead to a characteristic nonmonotonic beam-energy dependence of the kurtosis, $R_{4,2}^B = \chi_4^B/\chi_2^B$, of the net-baryon number distribution \cite{Stephanov:2008qz}. While this observable can straightforwardly be computed from the pressure via $\chi_n^B = T^{n-4}\frac{\partial^n p}{\partial\mu_B^n}$, there are many subtleties specific to heavy-ion collisions that prevent a direct comparison between first-principles QCD calculations and experiments \cite{Vovchenko:2023klh}. Still, basic qualitative features can be inferred directly from QCD. The QCD results from DSEs \cite{Lu:2025cls} and low-energy model results from the FRG \cite{Fu:2023lcm} on $R_{42}^B$ at chemical freeze-out  are confronted with the net-proton kurtosis measured by STAR \cite{STAR:2025zdq} in the left plot of Fig.\ \ref{fig:crit}. While functional methods find a nonmonotonic beam-energy dependence, no sign of critical scaling is seen along the freeze-out line. This behavior is a consequence of a sharp crossover transition, not necessarily a CEP. Still, its location is encoded in the height of the peak \cite{Fu:2023lcm}. In any case, 
the results show that data for $\sqrt{s} \approx 4\!-\!8$\,GeV will be crucial to make progress towards the experimental discovery of a chiral phase transition.

It is important the emphasize that the deviation from the noncritical baseline at $\sqrt{s} = 19.6$\,GeV (corresponding to $\mu_B \approx 200$\,MeV) reported in \cite{STAR:2025zdq} lies in a region of the phase diagram that is very well understood. Any relation between this deviation and a CEP is hence highly unlikely.

Since up and down quarks are very light, one may also ask if the chiral critical $O(4)$ scaling of QCD in the limit of massless up an down quarks has an observable effect \cite{Grossi:2021gqi}. For this to be possible, the chiral critical region needs to extend from zero to physical pion masses. The FRG, as a renormalization group technique, is ideally suited to address this question. A way to determine the size of the critical region is to assess where noncritical corrections to the scaling of the chiral condensate, $\langle \bar q q \rangle \sim m_\pi^{2/\delta} f_G(z)$, become relevant. It is shown in the right plot of Fig.\ \ref{fig:crit} that the critical exponent $\delta$ extracted from fits to the chiral condensate ceases to follow its universal value for $m_\pi \!\gtrsim\! 5$\,MeV at the chiral critical temperature $T_c \!=\! 142.58$\,MeV \cite{Braun:2023qak}. This implies that the critical region in mass-direction is very small. Even though its extent into the $T$-direction remains to be explored with the same rigor, chiral critical scaling seems to be unlikely for physical quark/pion masses.

\vspace{-5pt}
\section{The moat regime}\label{sec:moat}
\vspace{-5pt}

\begin{figure}[t]
\centering
\includegraphics[width=6.1cm,clip,valign=t]{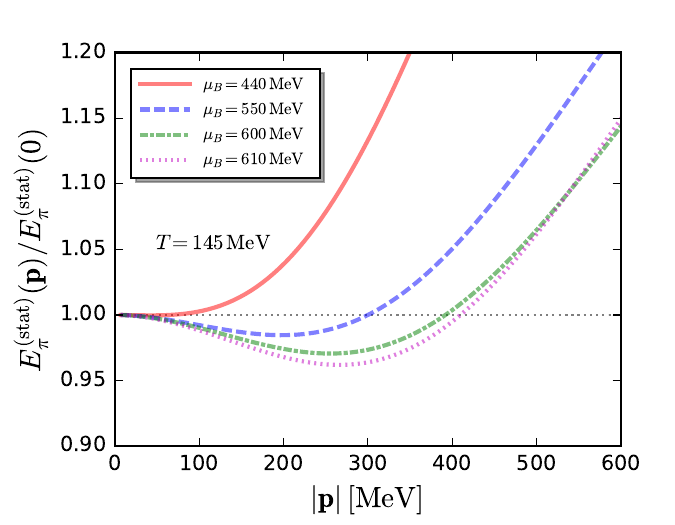}
\hfill
\includegraphics[width=6.7cm,clip,valign=t]{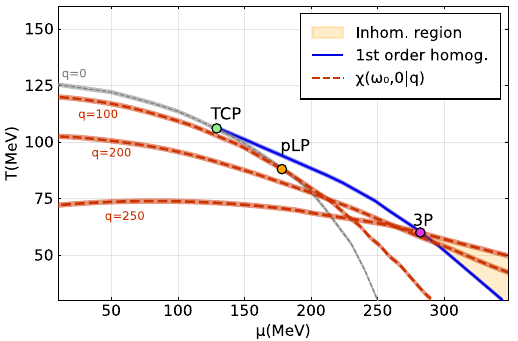}
\caption{\emph{Left:} Static pion energy just outside (solid red line) and inside the moat regime \cite{Fu:2024rto}, cf. Fig.\ \ref{fig:PD}.
\emph{Right:} Phase diagram of a rainbow-ladder QCD model in the chiral limit, featuring an inhomogeneous instability at small $T$ and large $\mu$ (yellow region) \cite{Motta:2024rvk}.}
\label{fig:moat} 
\vspace{-8pt}    
\end{figure}

Among the possible phases of dense QCD are those with spatial modulations. There are various possibilities ranging from actual inhomogeneous phases with long-range order, over liquid crystals with quasi--long-range order, to quantum pion liquids with no order at all \cite{Nussinov:2024erh}. The large number of possibilities complicates direct searches for such phases. The moat regime introduced in \cite{Pisarski:2021qof} is a superordinate feature common to all regimes with spatial modulations. 
It is hence a compelling target in the search for exotic phases. In the moat regime, the static energy of bosons is minimal at nonzero momentum. As shown in the blue region in Fig.\ \ref{fig:PD}, strong evidence for its existence in QCD was found with the FRG \cite{Fu:2019hdw}. In the left plot of Fig.\ \ref{fig:moat}, we show the static (zero-frequency) dispersion of pions as the moat regime is entered with increasing $\mu_B$ at fixed $T$ \cite{Fu:2024rto}. The location of the minimum determines the wavenumber of the underlying spatial modulation.

The moat regime is generated by spacelike particle-hole fluctuations of quarks at finite density, giving rise to a characteristic peak at nonzero momentum in the spectral function of pions and other mesons \cite{Fu:2024rto}. This can directly modify particle production at nonzero momentum and therefore lead to possible signals in heavy-ion collisions \cite{Pisarski:2021qof}, e.g., through Hanbury-Brown--Twiss correlations \cite{Rennecke:2023xhc} or dilepton yields \cite{Nussinov:2024erh}. The moat regime hence opens up the possibility for the experimental discovery of spatially modulated phases.

In the moat regime in Fig.\ \ref{fig:PD}, the static energy of mesons is always greater than zero. But it is possible that the energy becomes zero at the bottom of the moat somewhere in the unexplored region of Fig.\ \ref{fig:PD}. This would indicate an instability of the system towards the formation of an inhomogeneous phase. 
While an early DSE study assuming a very specific inhomogeneous phase \cite{Muller:2013tya} does not meet the benchmarks mentioned above, techniques to detect the more agnostic instabilities in QCD with state-of-the-art functional methods have recently been developed \cite{Motta:2023pks, Motta:2024agi, Motta:2024rvk, Fu:2024rto}. A first application to a rainbow-ladder QCD model is shown in the right plot of Fig.~\ref{fig:moat}. Together with numerous other model studies \cite{Buballa:2014tba}, this demonstrates that inhomogeneous phases are a serious possibility also in full QCD.

\vspace{-5pt}
\section{EoS of dense QCD matter}\label{sec:EoS}
\vspace{-5pt}

\begin{figure}[t]
\centering
\includegraphics[width=6cm,clip,valign=t]{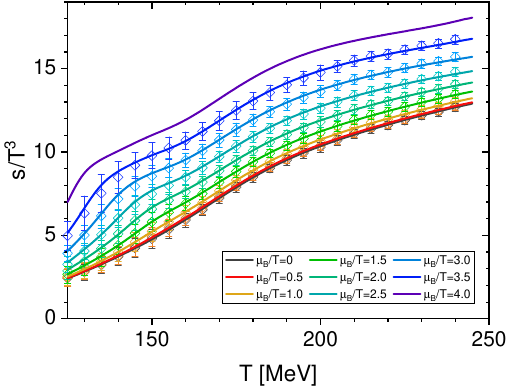}
\hfill
\includegraphics[width=6.8cm,clip,valign=t]{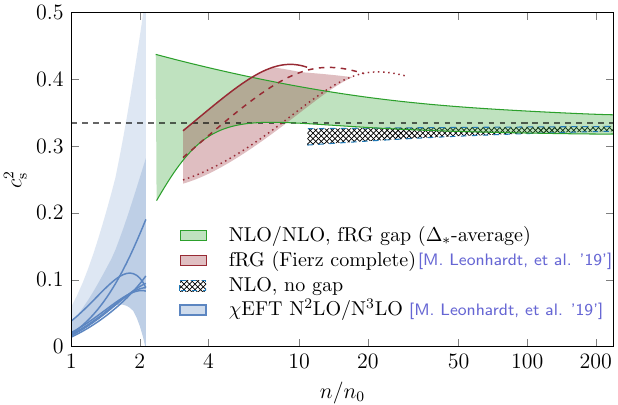}
\caption{\emph{Left:} Entropy density from functional QCD in comparison to lattice results at low and intermediate baryon density \cite{Lu:2025cls}. \emph{Right:} Speed of sound squared at large density in comparison to chiral effective field theory and weak-coupling results \cite{Leonhardt:2019fua, Braun:2021uua, Geissel:2024nmx}.}
\label{fig:eos} 
\vspace{-8pt}    
\end{figure}

We finally mention the rapid progress that functional methods have made in the determination of the QCD equation of state (EoS) at finite density. Starting from the first results in \cite{Isserstedt:2020qll}, systematic improvement have led to the determination of the EoS from first principles beyond the range of validity of lattice extrapolations, see Fig. \ref{fig:eos}. The left plot shows, on the example of the entropy density, that the EoS can by now be extended reliably to large $\mu_B/T$ using DSEs \cite{Lu:2025cls}. The important technical advancement from earlier calculations is the self-consistent treatment of the confining gluon background.

In addition, starting from exploratory DSE studies \cite{Muller:2016fdr}, the cold and dense, color-superconducting region of the phase diagram is being studied systematically using the FRG \cite{Braun:2019aow, Leonhardt:2019fua, Braun:2021uua}. A sizable two-flavor superconducting (2SC) diquark gap, $\Delta_{\rm 2SC} \approx 150-300$\,MeV, has been found at 5-10 times nuclear saturation density \cite{Leonhardt:2019fua, Braun:2021uua}. Note that existing bounds on the size of the gap based on astrophysical constraints only apply to color-flavor locked superconductors \cite{Kurkela:2024xfh, Geissel:2025vnp}. Interestingly, it has been demonstrated that the peak in the speed of sound, where the conformal value of $c_s^2 = 1/3$ is exceeded, is tied to the formation of the 2SC gap \cite{Leonhardt:2019fua}, see the right plot of Fig. \ref{fig:eos}.\\

\noindent
\textbf{Acknowledgements:} I thank the fQCD Collaboration \cite{fQCD} for discussions. This work is supported by the Deutsche Forschungsgemeinschaft (DFG, German Research Foundation) through the CRC-TR 211 ``Strong-interaction matter under extreme conditions'' -- project number 315477589 -- TRR 211.

%
%
%

\end{document}